\documentclass[reprint,
amsmath,amssymb,
aps,
]{revtex4-2}
\usepackage{graphicx}
\usepackage{dcolumn}

\begin{document}

\title{
Thermodynamic evidence for the formation of a Fulde-Ferrell-Larkin-Ovchinnikov phase
in the organic superconductor $\lambda$-(BETS)$_2$GaCl$_4$
}

\author{
S. Imajo$^{1,2,*}$,$\thanks{imajo@issp.u-tokyo.ac.jp}$
T. Kobayashi$^{3,4}$,
A. Kawamoto$^{3}$,
K. Kindo$^{2}$,
and
Y. Nakazawa$^{1}$
}
\affiliation{
$^1$Graduate School of Science, Osaka University, Toyonaka 560-0043, Japan\\
$^2$Institute for Solid State Physics, University of Tokyo, Kashiwa 277-8581, Japan\\
$^3$Graduate School of Science, Hokkaido University, Sapporo 060-0810, Japan\\
$^4$Present address: Graduate School of Science and Engineering, Saitama University, Saitama 338-8570, Japan
}

\date{\today}

\begin{abstract}
In this work, the thermodynamic properties of the organic superconductor $\lambda$-(BETS)$_2$GaCl$_4$ are investigated to study a high-field superconducting state known as the putative Fulde-Ferrell-Larkin-Ovchinnikov (FFLO) phase.
We observed a small thermodynamic anomaly in the field $H_{\rm FFLO}$ $\sim$ 10~T, which corresponds to the Pauli limiting field $H_{\rm P}$.
This anomaly probably originates from a transition from a uniform superconducting state to the FFLO state.
$H_{\rm FFLO}$ does not show a strong field-angular dependence due to a quasi-isotropic paramagnetic effect in $\lambda$-(BETS)$_2$GaCl$_4$.
The thermodynamic anomaly at $H_{\rm FFLO}$ is smeared out and low-temperature upper critical field $H_{\rm c2}$ changes significantly if fields are not parallel to the conducting plane even for a deviation of $\sim$0.5$^{\circ}$.
This behavior indicates that the high-field state is very unstable, as it is influenced by the strongly anisotropic orbital effect.
Our results are consistent with the theoretical predictions on the FFLO state, and show that the high-field superconductivity is probably an FFLO state in $\lambda$-(BETS)$_2$GaCl$_4$ from a thermodynamic point of view.
\end{abstract}

\maketitle
Spin-singlet superconductivity is suppressed by magnetic fields due to the orbital and paramagnetic pair-breaking effects.
The upper critical field of superconductivity, $H_{\rm c2}$, can assume different characteristics depending on whether the paramagnetic effect or the orbital effect predominates.
When the paramagnetic effect exceeds the orbital effect, $H_{\rm c2}$ is expected to correspond to the Pauli paramagnetic limit $H_{\rm P}$ at which the Zeeman splitting energy equals the superconducting gap energy.
In this case, $H_{\rm P}$ divides the low-temperature and high-field region of the $H$-$T$ phase diagram into a superconducting phase and a normal phase as a 1st-order phase boundary.
However, Fulde and Ferrell (FF)\cite{1} as well as Larkin and Ovchinnikov (LO)\cite{2} theorized the existence of unique pairing states above the $H_{\rm P}$ limit known as the FFLO state, in which a Cooper pair assumes a finite center-of-mass momentum on a Fermi surface split by the Zeeman effect.
This pairing reduces the spin polarization of Cooper pairs through the spatial modulation of a superconducting order parameter.
The observation of this unusual pairing state is constrained by two limitations, a clean electronic system\cite{3} and a large Maki parameter $\alpha_{\rm M}$\cite{4}.
This is because the mean free path of the electrons, $l$, is larger than the coherence length, $\xi$, in the regular spatial modulation and the FFLO state cannot be observed when $H_{\rm P}$ is lower than the orbital critical field $H_{\rm orb}$.
There are only a few superconductors that can satisfy these requirements.\cite{5,6,7,8,9,10,11,12,13,14,15,16,17,18,19,20,21}.
Layered organic superconductors are the best candidates to observe and thus study the FFLO state.
They are characterized by a small number of lattice defects and a two-dimensional (2D) superconductivity, which ensure a relatively large $\alpha_{\rm M}$ when fields are applied to the 2D plane.
Observation of the FFLO state has been reported in previous studies\cite{6,7,8,9,10,11,12,13,14,15,16,17,18,19}; however, the thermodynamic properties of the FFLO state are still unclear.
This is mainly due to experimental difficulties, such as the necessity to have strong magnetic fields and a precise control of the magnetic-field direction.
\begin{figure}
\begin{center}
\includegraphics[width=\hsize]{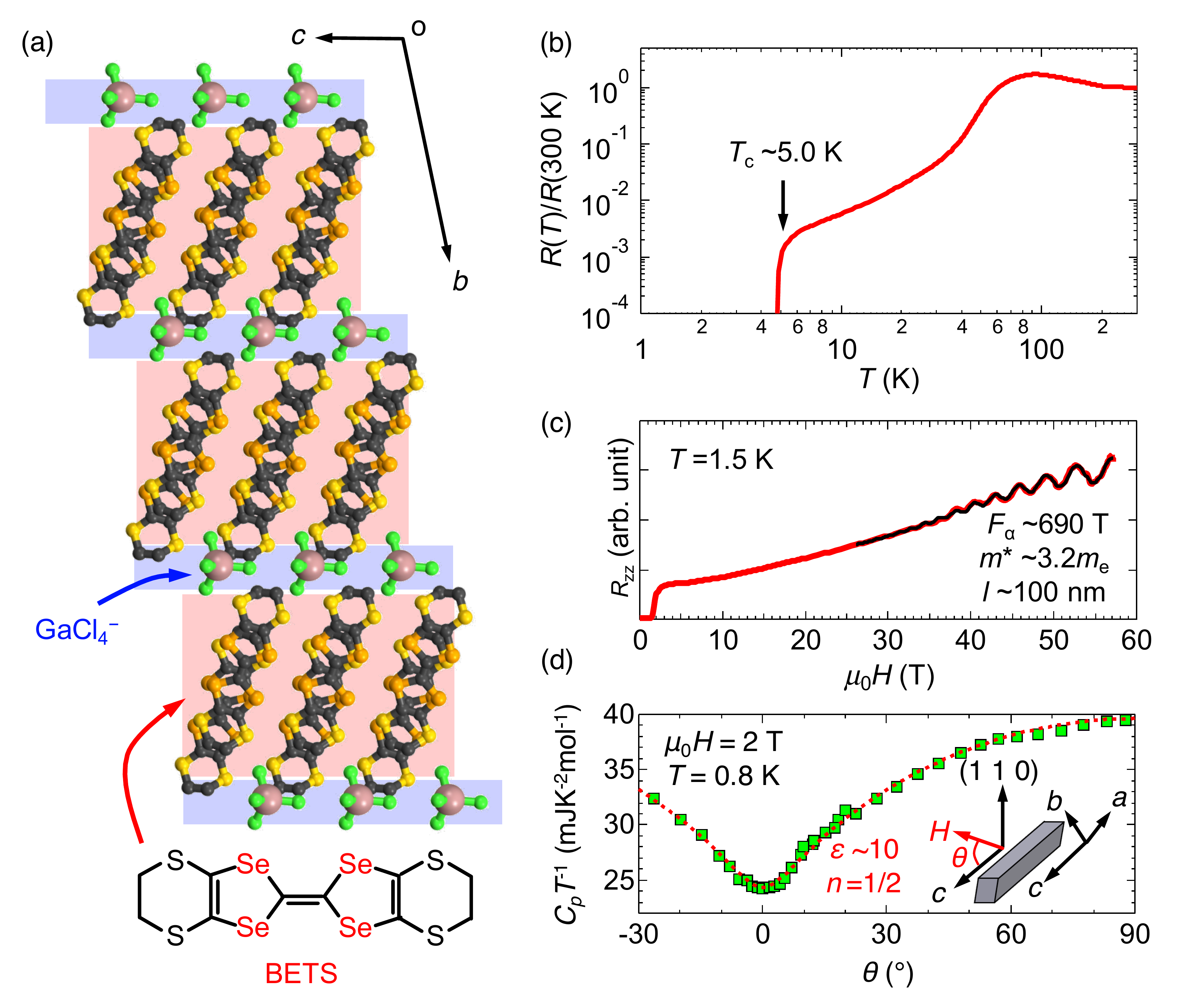}
\end{center}
\caption{
(a) Schematic view of the interlayer structure of $\lambda$-(BETS)$_2$GaCl$_4$, where BETS stands for bis(ethylenedithio)tetraselenafulvalene.
(b) Resistivity normalized at 300~K as a function of temperature.
The arrow indicates the superconducting transition temperature $T_{\rm c}$ at 5.0~K.
(c) Magnetic field dependence of the interlayer resistance at 1.5~K.
The black curve represents a fit obtained from the Lifshitz-Kosevich formula with an oscillation frequency $F_{\alpha}$ $\sim$ 690~T.
(d) Magnetic-field-angle dependence of the heat capacity at 2~T and 0.8~K.
The illustration depicts the definition of the angle $\theta$ from the $c$-axis.
The red dotted curve is obtained by fitting the data with Eq. (1).
}
\label{fig1}
\end{figure}

The organic superconductor $\lambda$-(BETS)$_2$GaCl$_4$, where BETS stands for bis(ethyleneditio)tetraselenafulvalene, is one of the candidate materials where an FFLO state can be expected.
As displayed in Fig.~\ref{fig1}a, this salt is composed of BETS and GaCl$_{4}^{-}$ in the ratio 2:1.
A quasi-2D electronic system composed of alternated organic and counter layers allows to stabilize the FFLO state by enhancing the Fermi surface nesting and suppressing the orbital effect when magnetic fields are applied parallel to the 2D conducting plane\cite{5,6,7,22,22p5}.
$\lambda$-(BETS)$_2$GaCl$_4$ exhibits a superconducting transition at $\sim$5~K\cite{8,11,14}, and the pairing state has a $d$-wave symmetry mediated by antiferromagnetic fluctuations\cite{23,24,25}.
Previous works\cite{8,11,14} reported that the BCS-FFLO transition occurs at about 10~T (=$H_{\rm FFLO}$) in a parallel field.
The transition was detected by observing changes in the dynamics and pinning of the vortices due to the FFLO modulation.
However, there is no thermodynamic evidence of the FFLO transition and these studies do not clearly state whether an FFLO state can be expected in the measured crystal.
Therefore, in this work we measured the heat capacity and the magnetocaloric effect (MCE) to detect this transition, and we described the thermodynamic properties of this putative FFLO state.

A single crystal measured in this experiment was electrochemically synthesized.
Using the usual ac-calorimetry and relaxation calorimetry techniques, the heat capacity was measured with a home-made high-resolution calorimeter\cite{26} in a $^3$He cryostat and with a 15~T superconducting magnet.
The MCE was measured using the same setup in fields swept at constant rates.
The field angle, controlled by a piezoelectric rotator, was monitored by the Hall voltage signal of a Hall device, mounted on a cell of the calorimeter\cite{25,26p5}.
The resolution of the determined angle in this measurement is about 0.1$^{\circ}$.
In addition, the electrical resistance of the crystal was also measured using a $^4$He cryostat put in a 60~T pulse magnet.

In Fig.~\ref{fig1}b and \ref{fig1}c we show the results of the electrical transport.
The temperature dependence of resistivity normalized at 300~K, $R$($T$)/$R$(300~K), shows a broad peak at 100~K and a superconducting transition at $T_{\rm c}$$\sim$5.0~K, which are in agreement with the previous study\cite{14}.
Figure~\ref{fig1}c presents the field dependence of resistance at 1.5~K.
Above 30~T, the magnetoresistance shows the Shubnikov de Haas (SdH) oscillations, which have a frequency of $\sim$690~T.
This result is also consistent with the data of the $\alpha$-orbit Fermi surface published in Ref.~\cite{27} and can be described by the Lifshitz-Kosevich formula (see the black curve).
From the analyses of the SdH oscillations, we obtain an effective mass ($m_{\rm eff}$) of 3.2$m_0$, which is comparable with 3.6$m_0$ evaluated in Ref.~\cite{27}.
Also, a mean free path ($l$) is determined as $\sim$100~nm, close to that of other FFLO candidates\cite{6,7}.
Figure~\ref{fig1}d shows the magnetic-field-angular dependence of the heat capacity at $T$=0.8~K and $\mu_{0}H$=2~T.
Previous studies\cite{23,25} have already reported data on the low-temperature heat capacity and discussed in detail the gap function.
$\theta$ is the angle between the magnetic field $H$ and the conductive layer, as shown in Fig.~\ref{fig1}d.
By changing the field direction, the perpendicular component of the field penetrates the conducting layer and strongly suppresses the superconductivity due to the orbital effect.
Based on the quasi-2D anisotropy of the superconductivity and the quasiparticle excitation, the angular dependence of $C_p$ can be formulated as the following formula\cite{26p5}:
\begin{equation}
\begin{split}
C_p(\theta)=C_p(0^{\circ})+\frac{\Delta C_p}{1-\epsilon^n}[1-({\rm cos}^2(\theta)+\epsilon^{2}{\rm sin}^2(\theta))^{n/2}],\\
(\Delta C_p=C_p(90^{\circ})-C_p(0^{\circ})).
\end{split}
\end{equation}
In this formula, $\epsilon$ and $n$ denote the anisotropy factor and the power index of the quasiparticle excitation, respectively.
Since the energy gap of the present superconductivity has line nodes on the Fermi surface\cite{23,24,25}, $n$ is 1/2.
In fact, the fit performed with $\epsilon$$\sim$10 and $n$=1/2 (see the red dotted curve) well reproduces the data while the fits performed using $n$=1 (fully gapped) do not provide plausible values of $\epsilon$.
Typically, $\epsilon$ is proportional to the square root of the ratio of the interlayer and in-plane effective masses $\sqrt{m_{\perp}/m_{\parallel}}$.
Therefore, $\epsilon$$\sim$10 ($m_{\perp}$/$m_{\parallel}$$\sim$100) indicates that the superconductivity has a quasi-2D characteristic\cite{28}.

\begin{figure}
\begin{center}
\includegraphics[width=\hsize]{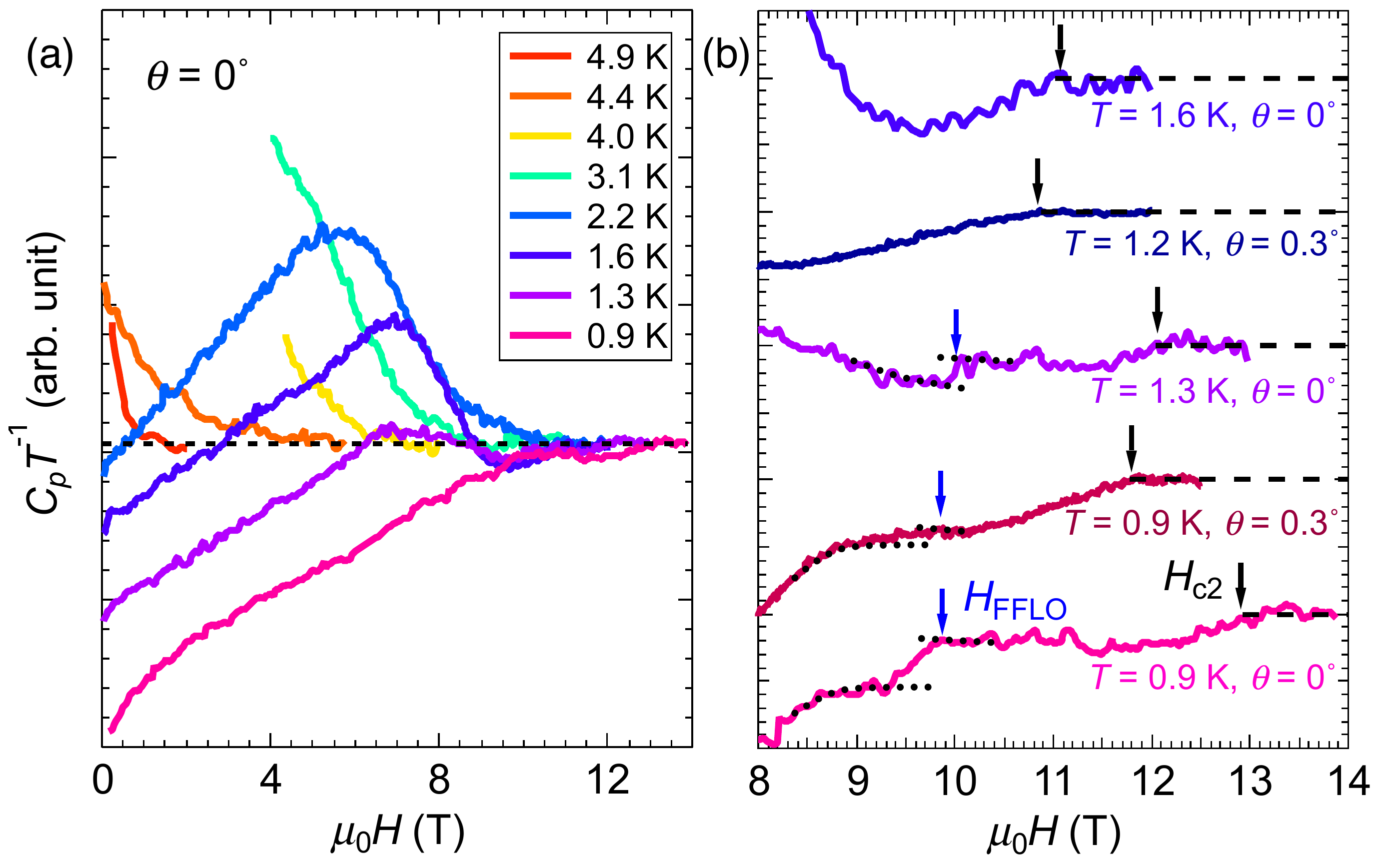}
\end{center}
\caption{
(a) $C_{p}T^{-1}$ vs $\mu_{0}H$ in a parallel magnetic field ($\theta$=0$^{\circ}$) at various temperatures (from 0.9~K to 4.9~K).
The dashed line represents the field-independent heat capacity of the normal state.
(b) A zoom of (a) in the low-temperature region above 8~T.
The dashed lines are plotted using different offsets to show the data clearly.
The black and blue arrows indicate $H_{\rm c2}$ and $H_{\rm FFLO}$, respectively.
The dotted curves are the guides for the eye to emphasize the step at $H_{\rm FFLO}$.
}
\label{fig2}
\end{figure}

Subsequently, we studied the thermodynamic characteristics of the high-field region near $H_{\rm c2}$.
The field dependence of the heat capacity at different temperatures is shown in Fig.~\ref{fig2}.
The dashed lines indicate the heat capacity of the normal state, and at each temperature the data merge with the corresponding dashed line at $H_{\rm c2}$.
The broad peak in Fig.~\ref{fig2}a reflects the field dependence of the quasiparticle density and the gap amplitude.
Since the behavior near $H_{\rm c2}$ at low temperatures is not clear from this plot, the results above 8~T are shown in Fig.~\ref{fig2}b.
In the same figure, we also show the results at $\theta$=0.3$^{\circ}$.
The black arrows indicate that $H_{\rm c2}$ strongly depends on $T$ and $\theta$.
In addition, the curve relative to $T$=0.9~K and $\theta$=0$^{\circ}$ shows a small bump at $\sim$10~T, that disappears at $T$=1.3~K and $\theta$=0.3$^{\circ}$.
Since this anomaly can be attributed to the BCS-FFLO transition at $H_{\rm FFLO}$ reported in Refs.~\cite{8,11,14}, the putative FFLO state appears in this field region above $H_{\rm FFLO}$ below $H_{\rm c2}$.
$H_{\rm FFLO}$ seems to have a small field-angle dependence.
This is reasonable since $H_{\rm FFLO}$ corresponds to $H_{\rm P}$ governed by the paramagnetic effect, which is almost isotropic due to the weak spin-orbit coupling in organic salts.
The fact that even a small misalignment can strongly reduce $H_{\rm c2}$ shows that the orbital effect can easily destabilize the FFLO state as suggested in Refs.~\cite{4,22p5,22p7}.
At 0.9 K, we observe the broad kink at 8--9~T, which can be detected as a small resistive jump\cite{14} and an anomaly of the penetration depth\cite{11}.
Although this anomaly is discussed in terms of a vortex transition, the detailed origin is still unclear.

\begin{figure}
\begin{center}
\includegraphics[width=\hsize,clip]{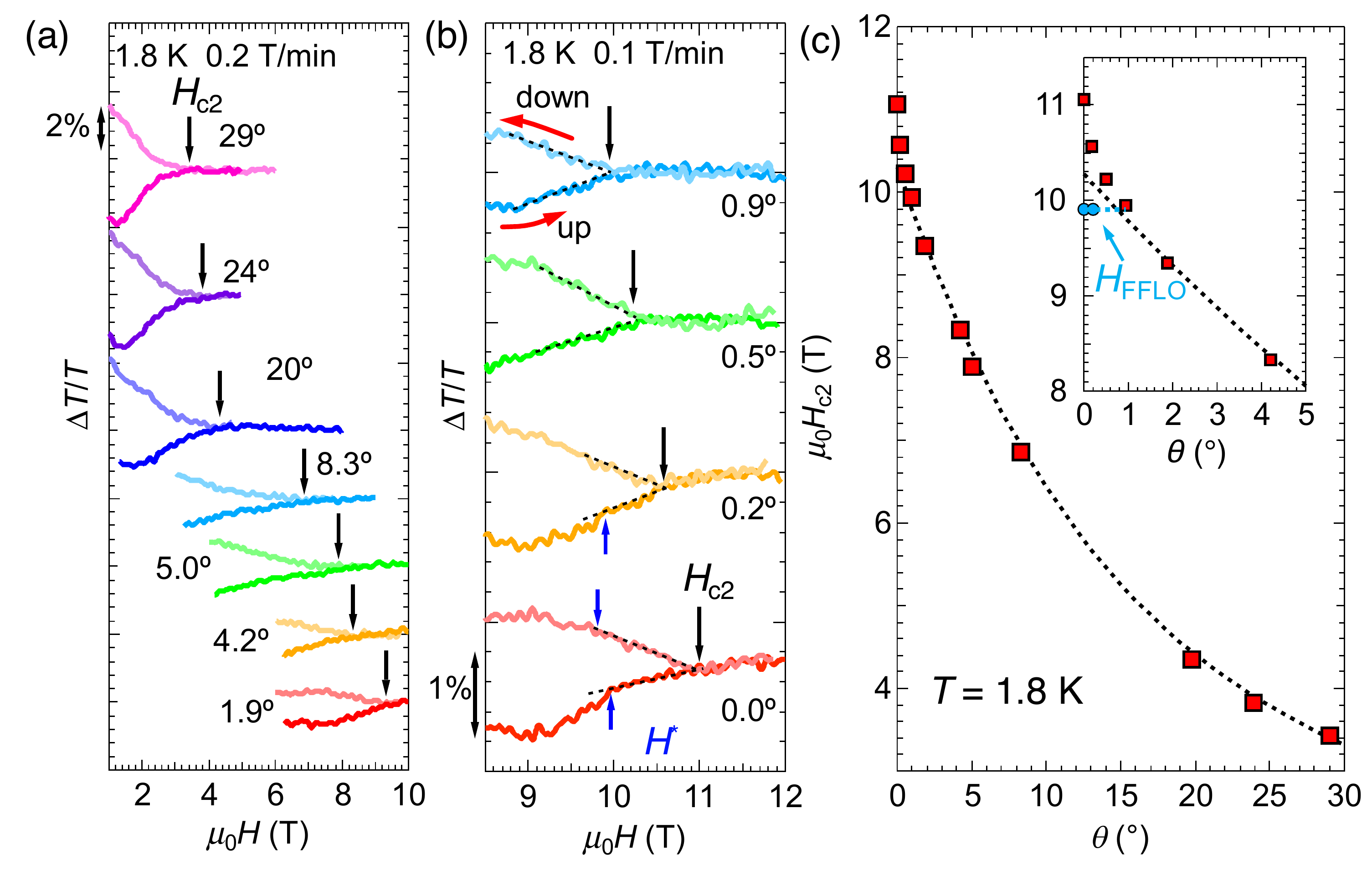}
\end{center}
\caption{
(a),(b) Relative temperature variation ($\Delta T$/$T$) at 1.8~K for different field angles.
The temperature variation is induced by the MCE that occurs at a field-sweep rate of 0.2~T/min (a) and 0.1~T/min (b).
The light-colored and the dark-colored curves represent the data obtained when the fields are swept up and swept down, respectively.
The black arrows indicate where the up-sweeps and down-sweeps separate ($H_{\rm c2}$) while the blue arrows indicate small steps at $H_{\rm FFLO}$$\sim$10~T.
(c) Angle dependence of $H_{\rm c2}$ derived from the data in (a) and (b).
The dotted curve is calculated using the 2D Tinkham model\cite{29}.
The inset shows a zoom on the small-angle region.
$H_{\rm FFLO}$ are also plotted as the blue dots.
}
\label{fig3}
\end{figure}

To further investigate the angle dependence of $H_{\rm c2}$ and $H_{\rm FFLO}$, we collected data on the MCE at 1.8~K.
In particular, we evaluated the relative variation of the sample temperature, $\Delta T$/$T$, when fields are swept up or down at the constant rates (see Fig.~\ref{fig3}).
The MCE is expressed as
\begin{equation}
\frac{\Delta T}{T}=-\frac{\tau}{T} \frac{d(\Delta T)}{dt}-\frac{1}{\kappa}\left( \frac{\partial S}{\partial H} \right)_{T}\dot{H}+\frac{\delta T}{T}
\end{equation}
, where $S$, $\tau$, $\kappa$, and $\delta T$ are the entropy, the thermal relaxation time, the thermal conductivity, and the irreversible heating due to dissipative losses, respectively.
We used magnetic-field sweep rates, $\dot{H}$, low ($\dot{H}$/$H_{\rm c2}$ is about 100 times smaller than $\tau$$^{-1}$) enough to make the system quasi-isothermal, $\Delta T$/$T$$\ll$1, and therefore, the first term in Eq.~2 is negligible.
Since the suppression of the superconductivity increases the entropy ($\partial$$S$/$\partial$$H$)$_{T}$$>$0, in the superconducting state $\Delta T$/$T$ is negative when $\dot{H}$$>$0 and positive when $\dot{H}$$<$0.
Since the entropy of the normal state does not strongly depend on field, ($\partial$$S$/$\partial$$H$)$_{T}$$\approx$0, $H_{\rm c2}$ is the point where the up-sweeps and down-sweeps separate (see the black arrows in Figs.~\ref{fig3}a and \ref{fig3}b).
In the parallel configuration (Fig.~\ref{fig3}b), a small step (the blue arrows) is observed at $\sim$10~T.
This is because the transition entropy is quite small at high fields near $H_{\rm c2}$ as for others FFLO candidates\cite{7,17,19,20}.
As we observed from the heat capacity data, this step is attenuated by slightly changing the inclination of the field.
Although the arrow for the data at 0.9~K in the down-sweep points the slight slope change, this anomaly is not clearly visible in the down-sweeps because the hysteresis and/or the dissipative losses covered the transition.
The asymmetric MCE depending on the sweeping direction implies that this anomaly is a 1st-order transition.
In fact, the BCS-FFLO is typically a 1st-order transition as the center-of-momentum discontinuously changes from zero to a finite value.
The symmetric MCE curves at $\theta$=0.9$^{\circ}$ indicate that the FFLO state disappears by the small tilt.
Figure~\ref{fig3}c shows the angle dependence of $H_{\rm c2}$ derived from the MCE data.
The dotted curve is calculated using the 2D-Tinkham model\cite{29}, which is used for the BCS-type superconductivity, with $H_{\rm c2}$(90$^{\circ}$)=1.8~T.
The plot shows that as $\theta$ increases $H_{\rm c2}$ decreases due to the strong orbital effect by the perpendicular-field component.
However,$H_{\rm c2}$ around 0$^{\circ}$ deviates from this model, as shown in the inset, this deviation starts when the field is about $H_{\rm FFLO}$.
This fact means that the superconductivity in the region $H$$>$$H_{\rm FFLO}$ and $\theta$$<$1$^{\circ}$ is not a BCS state, but a FFLO state.

\begin{figure}
\begin{center}
\includegraphics[width=\hsize,clip]{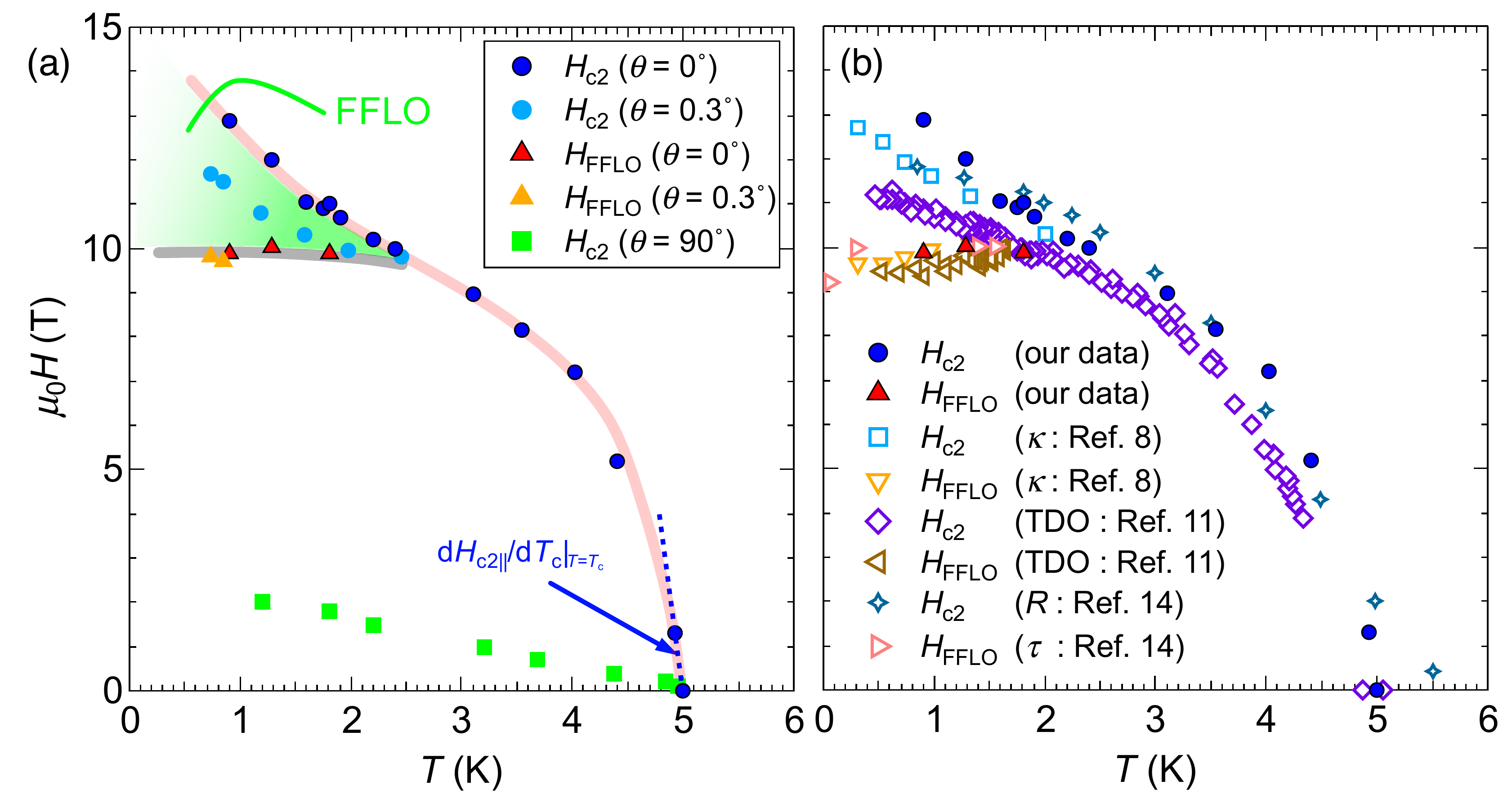}
\end{center}
\caption{
(a) Superconducting phase diagram of $\lambda$-(BETS)$_2$GaCl$_4$ obtained in our study.
The red and black translucent curves are the guides for the eye.
The blue dotted line represents (d$H_{\rm c2 \parallel}$/d$T_{\rm c}$)$|$$_{T=T_{\rm c}}$.
(b) Comparison between the data obtained in this work and those reported in the previous works\cite{8,11,14}.
}
\label{fig4}
\end{figure}

In Fig.~\ref{fig4}a, the $H_{\rm c2}$ and $H_{\rm FFLO}$ obtained in the present study are plotted in the $H$-$T$ phase diagram.
In the low-temperature and high-field region, the $H_{\rm c2}$ curve shows an upturn due to the formation of the FFLO state.
In Fig.~\ref{fig4}b, we compare the data collected at $\theta$=0$^{\circ}$ with those obtained in the thermal conductivity\cite{8}, rf-TDO\cite{11}, electrical transport\cite{14}, and magnetic torque measurements\cite{14}.
The low-temperature upturn is consistent with the theoretical predictions on the FFLO state in quasi-2D superconductors\cite{5,22,29p5,30}.
From the weak-coupling superconductivity relation, $H_{\rm P}$=$\Delta$/($\sqrt{2}$$\mu_{\rm B}$), $H_{\rm P}$ = 11.0~T if a gap amplitude $\Delta$$\sim$0.9~meV(=2.0$k_{\rm B}$$T_{\rm c}$) at 0 K\cite{23} is used.
This slight difference between $H_{\rm FFLO}$ and $H_{\rm P}$ can disappear if the refined formula $H_{\rm P}$=$\sqrt{2}$$\alpha$$k_{\rm B}$$T_{\rm c}$/$g^{\ast}\mu_{\rm B}$\cite{7,18,31} is used.
In this formula, the coupling strength $\alpha$\cite{32} and the effective $g$-value $g^{\ast}$, are also taken into account; this formula is valid for the other organic FFLO candidates as well\cite{6,7,16,18,33}.
In order to obtain $H_{\rm P}$=10~T, the renormalization factor $g^{\ast}$/$g$ has to be $\sim$1.1.
The $g^{\ast}$/$g$ factor of $\lambda$-(BETS)$_2$GaCl$_4$ has not been evaluated.
However, for organic salts the typical value of $g^{\ast}$/$g$ is in the range of 1.0-1.4\cite{31}.
Consequently, $H_{\rm P}$ =10~T is a reasonable result.
The relation $H_{\rm FFLO}$=$H_{\rm P}$ holds for all the organic FFLO candidates even though the fermiology and the superconducting pairing mechanisms might differ between materials.
This fact evidences that the BCS-FFLO transition in the organic salts is influenced only by the paramagnetic effect, while $H_{\rm c2}$ is influenced by also other factors, such as the orbital effect.

Our results are in good agreement with the theoretical prediction on the FFLO state.
Nevertheless, we should confirm that the crystal measured in this study can really show a FFLO state.
The FFLO state is possible only if the clean limit $l$$\gg$$\xi$ and the large Maki parameter $\alpha_{\rm M}$=$\sqrt{2}$$H_{\rm orb}$/$H_{\rm P}$ ($>$1.8) are met.
The in-plane $\xi_{\parallel}$ and $H_{\rm orb \parallel}$ can be determined from the superconducting phase diagram using the following formulas: $\xi_{\parallel}$=$\sqrt{\phi_0/2\pi H_{\rm c2}(90^{\circ})}$\cite{7,28} and $H_{\rm orb \parallel}$$\sim$$-$0.7$T_{\rm c}$(d$H_{\rm c2 \parallel}$/d$T$)$|$$_{T=T_{\rm c}}$\cite{34}, where $\phi_0$ is the magnetic flux quantum.
The corresponding values of $\xi_{\parallel}$ and $H_{\rm orb \parallel}$ are $\sim$14~nm at 2~K and $\sim$57~T, respectively.
Consequently, the $l$ of the present salt, $\sim$100~nm, is about 7 times higher than $\xi_{\parallel}$ and the Maki parameter $\alpha_{\rm M}$$_{\parallel}$$\sim$8 is much higher than 1.8, which is the minimum value of $\alpha_{\rm M}$$_{\parallel}$ that allows the formation of an FFLO state.
These values are sufficient to allow a BCS-FFLO transition at $H_{\rm P}$ and a FFLO state above $H_{\rm P}$.

In conclusion, we reported the results of the high-field thermodynamic measurements of the layered organic superconductor $\lambda$-(BETS)$_2$GaCl$_4$.
We detected the anomaly that originates from the BCS-FFLO transition in a parallel field at $\sim$10~T.
$H_{\rm FFLO}$ corresponds to $H_{\rm P}$ and does not have a significant field-angle dependence.
The formation of the FFLO state is only in a narrow region ($H$$>$10~T and $T$$<$2.5~K) and a slight deviation of the fields from the direction of the conductive layer strongly destabilizes the phase.
These results are exactly consistent with the characteristics expected for an FFLO state in 2D organic superconductors.
The results will help future studies in the characterization of the FFLO state.

We thank H. Kumagai (Hokkaido Univ.) for assistance in synthesizing the present salt.
This work was partially supported by Japan Society for the Promotion of Science KAKENHI Grant No. 20K14406.

\end{document}